\documentclass[final,3p,times,twocolumn]{elsarticle}
\journal{arXiv}

\usepackage{geometry}
\usepackage{amsmath} 
\usepackage{amssymb} 
\usepackage{braket} 
\usepackage{siunitx}

\usepackage{graphicx} 
\usepackage{booktabs} 
\usepackage{caption} 
\usepackage{float} 
\usepackage{comment}

\usepackage{lipsum} 

\usepackage{hyperref}
\hypersetup{
    colorlinks=true,
    linkcolor=blue,
    urlcolor=red,
    pdftitle={QuantumDNA},
    }

\usepackage{xcolor} 
\definecolor{codegreen}{rgb}{0,0.6,0}
\definecolor{codegray}{rgb}{0.5,0.5,0.5}
\definecolor{codepurple}{rgb}{0.58,0,0.82}
\definecolor{backcolour}{rgb}{0.95, 0.95, 0.95}

\usepackage{listing}
\usepackage{listings}

\lstdefinestyle{mystyle}{
    backgroundcolor=\color{backcolour},   
    commentstyle=\color{codegreen},
    keywordstyle=\color{magenta},
    numberstyle=\tiny\color{codegray},
    stringstyle=\color{codepurple},
    basicstyle=\ttfamily\footnotesize,
    breakatwhitespace=false,         
    breaklines=true,                 
    captionpos=b,                    
    keepspaces=true,                 
    numbers=left,                    
    numbersep=5pt,                  
    showspaces=false,                
    showstringspaces=false,
    showtabs=false,                  
    tabsize=2
}
\newcounter{bla}

\begin{document}


\begin{frontmatter}

\title{QuantumDNA: A Python Package for Analyzing Quantum Charge Dynamics in DNA and Exploring Its Biological Relevance}

\author[a]{Dennis Herb\corref{author}}
\author[b]{Marco Trenti}
\author[c]{Marilena Mantela}
\author[c]{Constantinos Simserides}
\author[a,d]{Joachim Ankerhold}
\author[a,d]{Mirko Rossini}

\cortext[author] {Corresponding author.\\ \textit{E-mail address:} dennis.herb@uni-ulm.de}
\address[a]{Institute for Complex Quantum Systems, Ulm University, 89081 Ulm, Germany}
\address[b]{Tensor AI Solutions GmbH, 89284 Pfaffenhofen an der Roth, Germany}
\address[c]{Department of Physics, National and Kapodistrian University of Athens, Panepistimiopolis, Zografos, GR-15784 Athens, Greece}
\address[d]{ Center for Integrated Quantum Science and Technology (IQST), Ulm-Stuttgart, Germany}

\begin{abstract}
The study of DNA charge dynamics is a highly interdisciplinary field that bridges physics, chemistry, biology, and medicine, and plays a critical role in processes such as DNA damage detection, protein-DNA interactions, and DNA-based nanotechnology. However, despite significant advances in each of these areas, knowledge often remains inaccessible to researchers in other scientific communities, limiting the broader impact of advances across disciplines. To bridge this gap, we present QuantumDNA, an open-source Python package for simulating DNA charge transfer (CT) and excited states using quantum physical methods. QuantumDNA combines an efficient Linear Combination of Atomic Orbitals (LCAO) approach with tight-binding (TB) models, incorporating open quantum systems techniques to account for environmental effects. This approach allows rapid yet accurate analysis of large DNA ensembles, enabling statistical studies of genetic and epigenetic phenomena. To ensure accessibility, the package features a graphical user interface (GUI), making it suitable for researchers across disciplines.
\end{abstract}


\end{frontmatter}




\section{Introduction} \label{sec:introduction}

DNA, the molecular blueprint of life, is not only known to encode most of the genetic information necessary to sustain life, but its complex molecular structure has long been studied as a platform for many physical phenomena. Among these, as early as sixty years ago, hypotheses were put forward for the possible relevance of quantum phenomena on the DNA molecule, both for engineering purposes and for a better understanding of the functioning of DNA in living cells \cite{Eley1962, Ladik1973}. For example, Löwdin's pioneering work demonstrated that classically forbidden proton tunneling through energy barriers could lead to DNA point mutations, shedding light on a novel mechanism of genetics \cite{Löwdin1963} and advancing our understanding of fundamental biological processes.

One area of research that has generated particular interest in the community is long-range CT \cite{Boon2002, Giese2002, Genereux2010} and long-lived excited states \cite{Crespo-Hernandez2004, Middleton2009, Schreier2015} on the DNA molecule. Understanding the charge dynamics in DNA is crucial for several scientific disciplines. From a physical and chemical perspective, these processes elucidate electron dynamics in complex macromolecular systems and inform the design of novel nature-inspired materials in molecular electronics \cite{Albuquerque2014, Wang2018}. Biologically, they are integral to genetic information processing and cellular functions, affecting DNA replication, repair, and transcription. In fact, characterization of the conducting properties of the DNA double helix has recently been explored as a novel approach to potentially uncover mechanisms of epigenetic regulation in living cells \cite{Siebert2023}. Medically, insights into these mechanisms may improve the diagnosis, treatment, and prevention of disease, particularly for conditions such as cancer \cite{Dandliker1997, Dandliker1998, Goodsell2001}.

To provide a computational framework for studying these processes, we present QuantumDNA, an open-source Python package specifically designed to simulate DNA charge dynamics using quantum physics methods. QuantumDNA uses a zoom-in/zoom-out approach to seamlessly integrate atomistic precision with system-level modeling. At the atomic level, charge interaction parameters are calculated using a built-in LCAO method amd serve as the basis for a coarser-grained TB model. This dual-scale approach ensures computational efficiency while maintaining the accuracy required for statistical analysis of large ensembles of DNA sequences, making QuantumDNA particularly suitable for statistical studies of genetic and epigenetic phenomena. Recently, the package was used to estimate exciton lifetimes and average dipole moments for all 16,384 possible seven-base pair DNA sequences \cite{Herb2024}, demonstrating its capability for high-throughput analysis of DNA electronic properties.

QuantumDNA integrates with publicly available biological databases such as the RCSB Protein Data Bank (PDB), allowing researchers to analyze a wide range of realistic DNA structures. Using molecular editing and visualization tools such as Biovia Discovery Studio and PyMol, users can introduce targeted mutations or modifications into DNA sequences and study their effects on CT and excited state properties. This functionality allows rapid screening of large DNA ensembles, facilitating the identification of sequences with significant electronic properties. These selected sequences can then be subjected to more detailed \emph{ab initio} calculations or used to guide experimental research. By combining physical modeling, structural biology, and computational chemistry, the package can serve as an interdisciplinary tool for studying DNA charge dynamics at multiple scales.

The manuscript is organized as follows. Section~\ref{sec:overview} provides a brief overview of the field of DNA physics and sets the stage for the following discussions. In Section~\ref{sec:package_structure}, we introduce the QuantumDNA package and outline its architecture. Section~\ref{sec:methodology} describes the theoretical framework and the computational methods used for the simulations. In Section~\ref{sec:features}, we demonstrate the capabilities of QuantumDNA by replicating the results of published research and presenting a biological application. Finally, Section~\ref{sec:conclusion} presents concluding remarks and discusses the limitations and planned future improvements of the package.


\section{ Charge and Energy Diffusion along DNA} \label{sec:overview}

\begin{figure*}[ht]
\centering
\includegraphics[width=1.\linewidth]{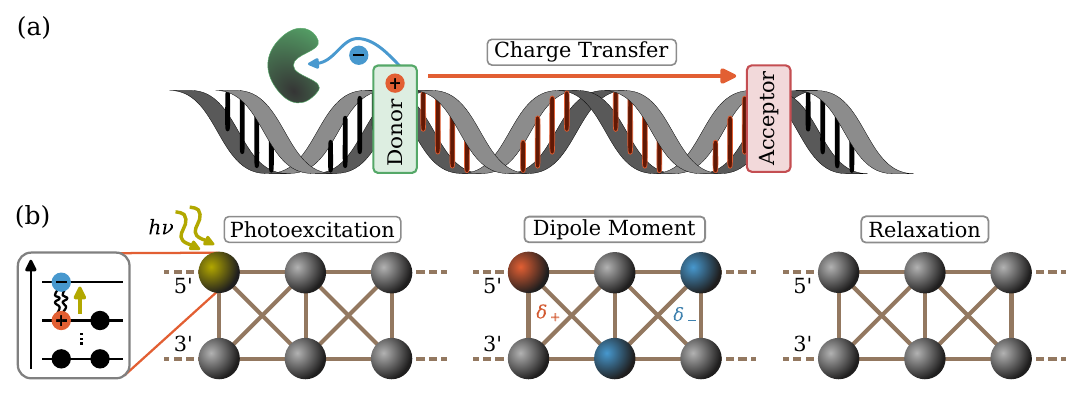}
\caption{\label{fig:1_CT_EET} 
    Schematic overview of the field of DNA-mediated coherent CT and DNA photophysics. 
    \textbf{(a)} A photoexcited redox molecule oxidizes a DNA base (donor), creating a hole (orange). The hole can move coherently along the DNA helix and oxidize a downstream acceptor molecule. Similarly, the reduction of a DNA base can produce an excess electron that follows the same migration mechanism.
    \textbf{(b)} (Left) Illumination of DNA produces an excitation by raising an electron from the HOMO to the LUMO level. The excited electron is bound to the remaining hole by an attractive Coulomb interaction, forming a Frenkel exciton. The DNA nucleobases forming the double helix are shown as spheres, with the top three spheres representing the top DNA strand and the bottom spheres representing the bottom DNA strand. (Middle) Due to the molecular orbital overlap between the DNA bases, the charges (electron and hole) as well as the exciton are transferred along the DNA double strand, indicated by the lines connecting the spheres. The electron and hole can separate and dissociate into a CT pair on an ultrafast timescale \cite{Bittner2006}, giving rise to an electric dipole moment. (Right) After a certain time, the DNA will convert the photoexcitation energy into thermal energy in a non-radiative manner \cite{Crespo-Hernandez2005}. }
\end{figure*}

\subsection{Charge Transfer}

Following the seminal characterization of the double-helical structure of DNA by Watson, Crick \cite{Watson1953} and Franklin, early hypotheses suggested that DNA might exhibit electrical conductivity, inspired by its structural similarity to the stacking patterns observed in one-dimensional molecular crystals \cite{Eley1962, Trixler2013}. Three decades ago, experiments confirmed that photoinduced CT can occur over long molecular distances between donor and acceptor molecules intercalated in DNA \cite{Murphy1993}. It is known that CT happens mainly through the pi-stacked DNA nucleobases and is very sensitive to changes in the stacking. Therefore, it has been proposed as a signaling and sensing mechanism that may allow DNA-binding proteins to detect and repair mutations and defects in the DNA sequence, as suggested by Barton et al. \cite{Arnold2016, Zwang2018}.

Over short distances of 3-4 base pairs quantum mechanical tunneling (superexchange) is assumed to dominate CT \cite{Giese2001}. 

DNA nucleobases can undergo redox reactions resulting in the insertion of a hole (oxidation) or an electron (reduction) by external stimuli, represented by the green object in Fig.~\ref{fig:1_CT_EET}(a). The specific mechanisms driving these events vary with experimental conditions \cite{Genereux2010, Zwang2018}, but the focus here is on the resulting charge state of the nucleobase, which acts as the initial charge donor. Once initiated, the charge can propagate along the DNA chain and reach a distant acceptor. 

\subsection{Excited States}

Photoexcited energy transfer is a fundamental process in nature and the basis for life on Earth. In DNA, the absorption of ultraviolet radiation (UVB) from sources such as sunlight or lasers has been extensively studied both experimentally and theoretically \cite{Crespo-Hernandez2004, Middleton2009, Schreier2015}. The remarkable photochemical stability of DNA is due to its ability to dissipate photoexcitation energy as thermal energy within picoseconds through non-radiative processes. This rapid energy conversion helps to prevent UV-induced mutations.

Nevertheless, transient absorption and pump-probe experiments by Kohler et al. \cite{Crespo-Hernandez2005, Takaya2008} indicate that the excited state lifetime is increased in stacked DNA bases compared to DNA monomers. This could explain how photomutagenic damage to DNA occurs, for example, cyclobutane pyrimidine dimers (CPDs) linking neighboring T-bases, which are a major contributor to skin cancer.

A physical model of DNA photoexcitation is shown in Fig.~\ref{fig:1_CT_EET}(b). Energy absorption promotes an electron from the highest occupied molecular orbital (HOMO) to the lowest unoccupied molecular orbital (LUMO). As shown on the left side of Fig.~\ref{fig:1_CT_EET}(b), this creates an excited state in which the electron and the remaining hole are bound by a Coulomb attraction, forming a Frenkel exciton. The electron and hole can move along the DNA double strand by molecular orbital overlap between adjacent bases. On an ultrafast timescale, they can separate to form a charge-separated state, generating an electric dipole moment, as shown in the centre of Fig.~\ref{fig:1_CT_EET}(b). Energy dissipation via an internal conversion mechanism relaxes the DNA molecule to its equilibrium state. 

\section{QuantumDNA Package Structure} \label{sec:package_structure}

\begin{figure}[H]
\centering
\includegraphics[width=1.\linewidth]{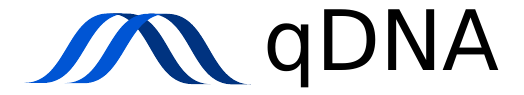}
\caption{\label{fig:0_qDNA_logo} 
    The QuantumDNA logo. The blue waves
symbolise the wave-like behaviour of particles inherent in
quantum mechanics, while also representing the iconic
structure of the DNA double helix.}
\end{figure}

\subsection{Workflow Structure}

\begin{figure}[ht]
\centering
\includegraphics[width=1.\linewidth]{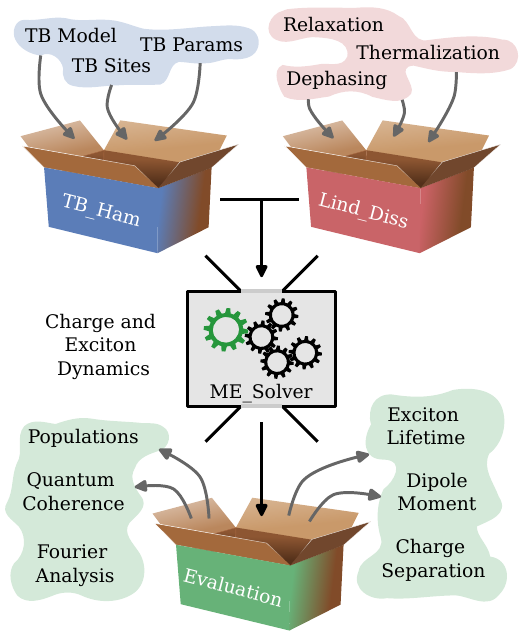}
\caption{\label{fig:2_qDNA_structure} Schematic overview of the structure of the QuantumDNA package. 
(Blue) A TB model is selected and tuned to generate the TB Hamiltonian describing the isolated DNA system via the class \texttt{TB\_Ham}. 
(Red) Lindblad operators are introduced to account for system-bath interactions induced by the DNA environment, including electron-hole recombination, via the class \texttt{Lind\_Diss}. 
(Grey) The TB Hamiltonian and Lindblad operators are combined into a master equation to calculate the charge dynamics along the DNA segment within the class \texttt{ME\_Solver}. 
(Green) The package includes a set of analysis tools to compute and visualize results, providing multiple observables to evaluate and interpret DNA charge dynamics. }
\end{figure}

The QuantumDNA package can be easily installed using the Python pip installer with the following command: 

\begin{lstlisting}[language=bash]
pip install qDNA
\end{lstlisting}

The study of DNA charge dynamics in a coarse-grained quantum framework requires addressing different challenges at different scales. To ensure clarity, modularity, and ease of future development, each class in QuantumDNA has been designed to handle a specific aspect of the simulation. Two particularly important classes, illustrated in Fig.~\ref{fig:2_qDNA_structure}, focus on incorporating environmental interactions, which are essential because DNA in vivo is always embedded in a complex cellular environment. These core components include the isolated DNA system (\texttt{TB\_Ham}, blue) and its environmental interactions (\texttt{Lind\_Diss}, red), following the well-established Lindblad formalism for modeling open quantum systems. A complete overview of all QuantumDNA classes is given in Tab.~\ref{tab:classes} in the Appendix. Meanwhile, Fig.~\ref{fig:2_qDNA_structure} presents a conceptual illustration of the overall structure of the package.


The QuantumDNA package aims to bridge the gap between experimental DNA measurements and tangible predictions about the consequences of charge dynamics along the DNA molecule. Therefore, the first step is to evaluate suitable parameters for the TB model from the measured distribution of atoms of the DNA sample of interest. The input parameters to define the DNA TB model can be provided in three different ways: (1) by direct calculation within the package using the state-of-the-art LCAO methods developed by Mantela et al. \cite{Mantela2021} (see Section~\ref{sec:methodology}) provided the geometric distribution of atoms in the DNA segment; (2) by using parameters from the literature, which are already included and available in the package and summarized in Table~\ref{tab:tb_params}; (3) by providing a custom set of parameters as input by the user. These atomistic parameters form the basis for a coarser-grained TB framework, allowing users to model the physical properties of DNA at larger scales. 

When calculating the parameters from the distribution of atoms in the system of interest, this distribution can be provided in XYZ or PDB format, the standard file format used by the Protein Data Bank (PDB) to represent DNA structures. The preparation of the molecular structure file includes the removal of the backbone atoms, as the influence of the backbone is taken into account indirectly in the parameterization of the environmental effects, as discussed below. Further developments currently underway will implement the ability to investigate possible CT phenomena directly mediated by the backbone to further refine the accuracy of the model, as recently investigated by Kordas et al. This pre-processing step can easily be done manually or by using freely available software such as Biovia Discovery Studio \cite{BIOVIA2016} or similar molecular editing tools.

Given a specific DNA segment, the preferred strategy for obtaining the required parameters includes the choice of the best TB model for the analysis to be performed. The choice of the TB model (to be selected from several built-in options as shown in Fig.~\ref{fig:appendix_TB}) is crucial as it determines the trade-off between the required level of detail and computational efficiency. Fig.~\ref{fig:appendix_scaling} shows an example of the scaling of computational times as a function of the chosen TB model and the length of the DNA sequence. Once the DNA sequence of interest and the specific TB model have been chosen, the isolated system is well-defined. 

In the following example, we set up a model to simulate the dynamics of an exciton initially localized at the leftmost guanine (G) within the GCG sequence occupying the 2nd, 3rd, and 4th base pairs of the 1BNA dodecamer. The 1BNA structure is widely recognized as a benchmark in computational structural biology due to its historical significance as one of the first high-resolution B-DNA structures \cite{Drew1981}. The corresponding PDB file can be obtained from \verb|rcsb.org|.

For this simulation, we use the Extended Ladder Model (ELM) shown in Fig.~\ref{fig:4_TB}(b) and parameterize it using two different approaches: (1) computing parameters from a PDB file that provides atomic-level structural information, and (2) using a predefined parameter set from the literature.

To compute TB parameters for the 1BNA molecular structure, we select the PDB file and the desired TB model, and proceed with calculating and saving the parameters:

\begin{lstlisting}[language=Python]
from qDNA import *

# Select TB model and calculate TB params
tb_model = 'ELM'
convert_pdb_to_xyz("1BNA.pdb")
HOMO_dict, LUMO_dict = calc_tb_params(["1BNA"], tb_model)

# Save the calculated TB params
wrap_save_tb_params(HOMO_dict, "1BNA", "hole", tb_model, unit="meV")
wrap_save_tb_params(LUMO_dict, "1BNA", "electron", tb_model, unit="meV")
\end{lstlisting}

\noindent
Once the set of parameters has been generated, the DNA model Hamiltonian can be easily constructed using the \texttt{TB\_Ham} class. However, as mentioned above, QuantumDNA's capabilities extend beyond the simulation of isolated TB models. To account for environmental effects, the \texttt{Lindblad\_Diss} class allows the inclusion of various dissipative processes, allowing users to model system-bath interactions relevant to their specific research questions. This functionality makes it possible to introduce quantum decoherence and relaxation effects, or to simulate the escape of excitons into the environment - phenomena that have been experimentally investigated, for example, by Kohler et al. \cite{Crespo-Hernandez2005}.

In this example, we restrict the system-bath interactions to those governing electron-hole recombination and energy dissipation into the thermal environment:

\begin{listing}
\begin{lstlisting}[language=Python]
# Select the DNA segment of interest
upper_strand = ['02G', '03C', '04G']
lower_strand = ['23C', '22G', '21C']

# Choose options
kwargs = dict(unit = "rad/ps", 
              relax_rate = 3.,
              source = "1BNA",
              lower_strand=lower_strand))
\end{lstlisting}
\end{listing}

In the following, we use the built-in set of parameters presented in Ref.~\cite{Hawke2010} with the source identifier \texttt{Hawke2010}:

\begin{listing}
\begin{lstlisting}[language=Python]
# Select the DNA segment of interest
upper_strand = ['G', 'C', 'G']
lower_strand = ['C', 'G', 'C']

# Choose options
kwargs = dict(unit = "rad/ps", 
              relax_rate = 3.,
              source = "Hawke2010")
\end{lstlisting}
\end{listing}

Once the parameters are set, the following code creates the instances for the TB Hamiltonian of the isolated DNA double-strand (\texttt{TB\_Ham} class) and its interaction with the environment (\texttt{Lindblad\_Diss} class):

\begin{lstlisting}[language=Python]
# Create instances of TB_Ham and Lindblad_Diss
dna_seq = DNA_Seq(upper_strand, tb_model, lower_strand=lower_strand)
tb_ham = TB_Ham(dna_seq, **kwargs)
lindblad_diss = Lindblad_Diss(tb_ham, **kwargs)
\end{lstlisting}

The system dynamics and various observables can now be analyzed using the \texttt{ME\_Solver} class, which uses the QuTiP master equation solver \cite{qutip1, qutip2}. This class allows the calculation of key properties, including the population dynamics and charge coherences within the modeled DNA segment. In addition, several built-in observables are available, such as the average excitonic lifetime and the dipole moment resulting from electron-hole separation. In the following, we demonstrate how to compute and visualize the time evolution of electrons, holes and excitons, as well as how to evaluate the aforementioned observables:

\begin{listing}
\begin{lstlisting}[language=Python]
me_solver = ME_Solver(tb_ham, lindblad_diss, **kwargs)

# Plot population dynamics
fig, ax = plot_pops_heatmap(me_solver)

# Calculate exciton lifetime
lifetime = calc_lifetime (upper_strand, tb_model, **kwargs)
print (f"Exciton Lifetime {lifetime} fs")

# Calculate dipole moment 
dipole = calc_dipole (upper_strand, tb_model, **kwargs)
print (f"Charge Separation {dipole} A")
\end{lstlisting}
\end{listing}

A complete list of the default settings used in the classes can be found in Tab.~\ref{tab:defaults}, while a detailed description of all available classes and functions can be found in the official \href{quantumdna.readthedocs.io/en/latest/}{QuantumDNA documentation webpage}.

\subsection{GUI Setup and Usage}

\begin{figure*}[ht]
\centering
\includegraphics[width=1.\linewidth]{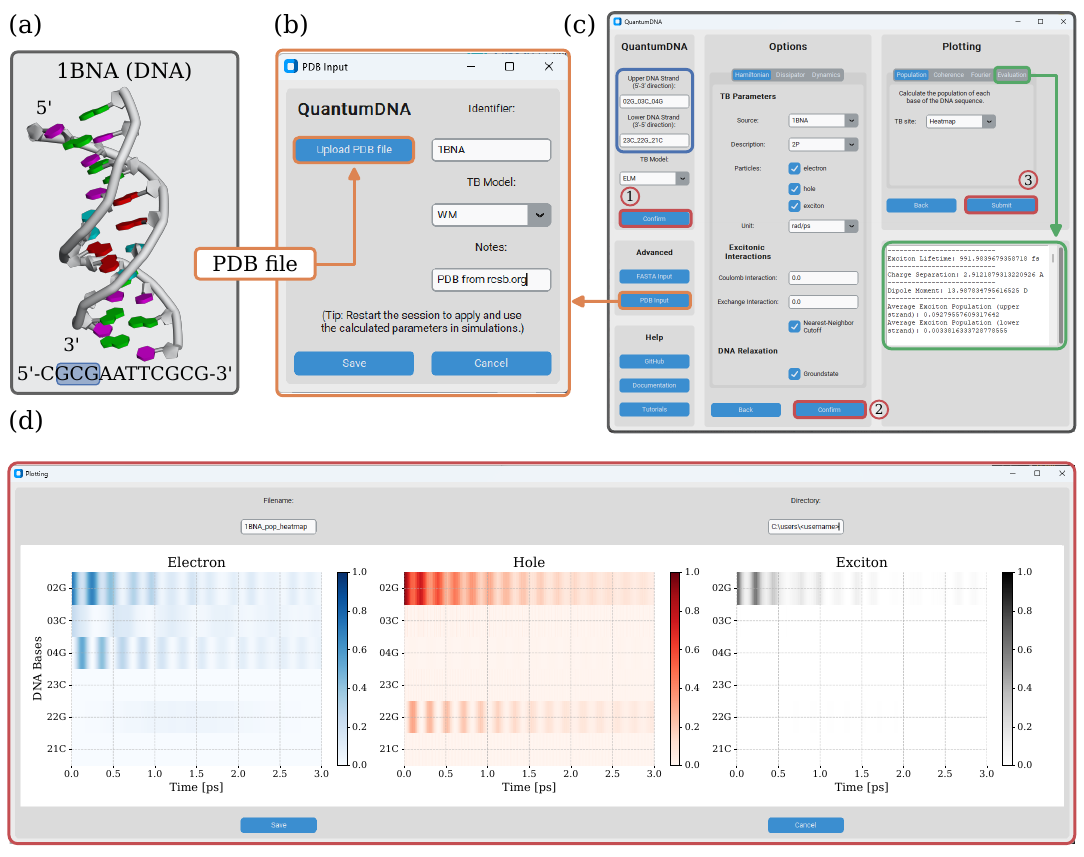}
\caption{\label{fig:5_GUI} 
Quantum-physical simulations with real DNA geometries via the GUI.
\textbf{(a)} A PDB file containing the DNA geometry was obtained from \texttt{rcsb.org} (identifier: \texttt{1BNA}) and modified using Biovia Discovery Studio \cite{BIOVIA2016} by removing the sugar-phosphate backbone. The subsequence selected for simulation is highlighted in blue.
\textbf{(b)} The PDB Input window of the GUI allows the user to upload the modified PDB file, specify an identifier, and select a TB model. Clicking the "Save" button calculates TB parameters tailored to the DNA geometry.
\textbf{(c)} To simulate the highlighted sequence from (a), set the upper strand to \texttt{02G\_03C\_04G} and the lower strand to \texttt{23C\_22G\_21C}. Make sure that the identifier (e.g. \texttt{1BNA}) is selected as the source. Exciton calculations can be performed using the Evaluation tab, with the results displayed in the console at the bottom right (highlighted in green).
\textbf{(d)} The Plot window provides a heatmap visualization of the time-evolved populations for the DNA sequence highlighted in (a). All simulation steps can be performed programmatically without the GUI, for example using Jupyter notebooks.}
\end{figure*}

In addition to its Python-based implementation, the QuantumDNA package features an intuitive and user-friendly GUI designed to facilitate its use by researchers with limited experience in computational techniques. To demonstrate its functionality, we present an example workflow based on the previously introduced 1BNA dodecamer. A schematic representation of the 1BNA structure is shown in Fig.~\ref{fig:5_GUI}(a).

Once the PDB file has been processed, the GUI can be started using the following command, either from a terminal or within a Jupyter notebook:

\begin{listing}
\begin{lstlisting}[language=Python]
from qDNA.gui import qDNA_app

app = qDNA_app()
app.mainloop()

\end{lstlisting}
\end{listing}

The main GUI interface, shown in fig.~\ref{fig:5_GUI}(c), provides an intuitive environment for setting parameters and visualizing results efficiently. To generate a new set of TB parameters specific to the 1BNA dodecamer, select the "PDB Input" option (highlighted in orange). This action opens a separate window, as shown in Fig.~\ref{fig:5_GUI}(b). Upload the edited PDB file, assign an identifier, and select a TB model. Once the TB parameters have been calculated, restart the GUI to ensure that they are correctly loaded into the main menu.

In the menu, specify the DNA segment to be simulated (in this example, GCG, highlighted in fig.~\ref{fig:5_GUI}(a)) by entering it in the appropriate field (highlighted in blue in Fig.~\ref{fig:5_GUI}(c)) and confirming with the first button (red 1). Additional simulation settings can then be adjusted, the default values are listed in the Tab.~\ref{tab:defaults}. It is important to update the source field to \texttt{1BNA} to ensure that the newly generated TB parameters are used instead of the default \texttt{Hawke2010} settings. After configuring all options, press the second confirmation button (red 2), which will initialize an instance of the \texttt{ME\_Solver} class in the background.

The right panel of the interface provides various calculation and visualization options. The excited state properties can be calculated (green frame), while the charge dynamics can be visualized using the heatmap option, which provides an intuitive representation of the charge dynamics similar to \cite{Bittner2007}. Finally, pressing the Submit button (red 3) generates the heatmap, which can be saved for further analysis, as shown in Fig.~\ref{fig:5_GUI}(d).

\section{Methodology} \label{sec:methodology}

\subsection{Linear Combination of Atomic Orbitals (LCAO)}

Ab initio methods, such as Density Functional Theory (DFT) and post-Hartree-Fock (post-HF) techniques, provide accurate electronic structure calculations for small DNA fragments near equilibrium. More advanced approaches, such as Real-Time Time-Dependent DFT (RT-TDDFT), enable the study of dynamical, out-of-equilibrium phenomena. However, the computational cost of these methods scales rapidly, making them impractical for systems beyond 3–4 nucleotides, thereby limiting their applicability to long-range CT studies. To efficiently analyze large ensembles of DNA sequences, alternative methods with higher numerical efficiency are required.

A significant breakthrough came with the LCAO-based approach introduced by Endres, Cox, and Singh \cite{Endres2002}, which calculates CT parameters by deriving molecular orbitals (MOs) from the overlap of atomic orbitals (AOs) using the semi-empirical Slater-Koster method. The open parameters in this model were optimized to reproduce results from high-accuracy ab initio calculations. Building on this foundation, Hawke et al. \cite{Hawke2010} refined the parameterization to generate a more comprehensive set of TB parameters at both the single-base and base-pair levels. More recently, Mantela et al. \cite{Mantela2021} further extended the LCAO approach by incorporating all valence orbitals, moving beyond Hückel-type LCAO. This improvement enhanced its applicability to DNA charge transfer and transport studies, including cases involving distorted or mutated DNA sequences \cite{MLS:2023}.

To achieve the required accuracy for charge dynamics simulations while maintaining computational efficiency, QuantumDNA employs an LCAO method to compute CT parameters between neighboring bases in the TB model (DNA bases/base pairs). Each MO is expressed as a linear combination of AOs:

\begin{align} 
\ket{\text{MO}} = \sum_{\alpha=1}^N \sum_{i=1}^I c_{i \alpha} \ket{\phi_{i \alpha}},
\end{align}

\noindent
where $i\in I$ runs over all the AOs of a given atom, and $\alpha\in N$ runs over all atoms in the molecule. 

Following the current state-of-the-art, we include in our calculations all valence orbitals of the atoms constituting each DNA base, including the 2s, 2p$_x$, 2p$_y$, and 2p$_z$ orbitals for the carbon (C), nitrogen (N), and oxygen (O) atoms, and the 1s orbital for the hydrogen (H) atoms. The AOs are assumed to be orthogonal to each other, i.e. $\braket{\phi_{i\alpha} | \phi_{j\beta}} = \delta_{ij}\delta_{\alpha\beta}$.

The MOs are obtained as eigenstates of the molecular Hamiltonian with the their energies being the corresponding eigenenergies. Therefore, we determine the coefficients $c_{i \alpha}$ by diagonalizing the molecular Hamiltonian that is constructed from the constituent AOs. To quantify the AO overlap $J_\chi$, we employ Slater-Koster two-center transfer integrals, where $\chi\in\{ss,sx,xx,xy\}$ represents the overlapping orbitals \cite{Slater1954}. 

\begin{align*} 
J_{\mathrm{ss}} & = V_{\mathrm{ss} \sigma} \\
J_{\mathrm{sx}} & = \xi_x V_{\mathrm{sp} \sigma} \\
J_{\mathrm{xx}} & = \xi_x^2 V_{\mathrm{pp} \sigma} + \left(1-\xi_x^2\right) V_{\mathrm{pp} \pi} \\
J_{\mathrm{xy}} & = \xi_x \xi_y \left(V_{\mathrm{pp} \sigma} - V_{\mathrm{pp} \pi}\right). 
\end{align*}

\begin{figure}[ht]
\centering
\includegraphics[width=1.\linewidth]{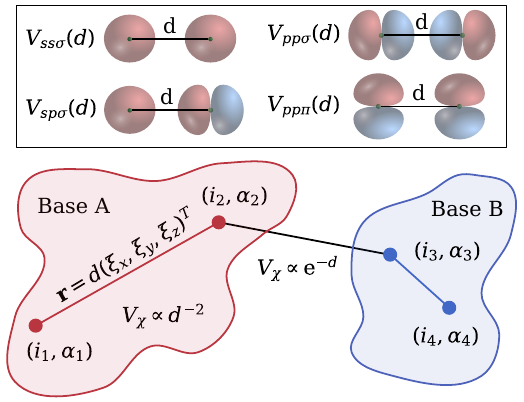}
\caption{\label{fig:3_LCAO} 
   Illustration of the calculation of Slater-Koster two-center transfer integrals for the LCAO method. The tuples represent the $\alpha$ AO for the $i^{th}$ atom. The intrabase AO overlap decreases quadratically with the two-center distance $d$ as described by Harrison \cite{Harrison1989, Harrison1999}. The directional cosines $\xi$ required for the Slater-Koster expressions are calculated from the vector connecting each intrabase atom. For distances exceeding typical covalent bond lengths, such as those encountered when evaluating interbase AO overlap, we use modified Harrison expressions with exponential decay as suggested by \cite{Hawke2010, Mantela2021}. \textit{(Box)} Overview of the possible AO interactions between $s$ and $p$ orbitals. The strength of the interaction is measured by $V_\chi$ and depends on the two-center distance $d$ between the AOs.}
\end{figure}

The overlap of AOs depends on both the distance $d$ between the atoms and the angles integrated into the directional cosines $\xi$ derived from the vector connecting the atoms as shown in Fig.~\ref{fig:3_LCAO}.  

To calculate each interaction value $V_{\chi}$ we use the quadratic decay in Eq.~\eqref{eq:quadratic} as proposed by Harrison \cite{Harrison1989, Harrison1999}, which is valid for the typical interatomic distances within a DNA base. However, since these expressions are limited to intrabase interactions, we include an exponentially decaying term in Eq.~\eqref{eq:exponential} to account for the overlap between AOs of neighboring DNA bases \cite{Hawke2010, Mantela2021}.

\begin{align} 
V_\chi^{\text{intra}} & = C_\chi \frac{\hbar^2}{m d^2} \label{eq:quadratic}\\
V_\chi^{\text{inter}} & = C_\chi \frac{\hbar^2}{m d_0^2} e^{-\frac{2}{d_0} \left(d-d_0\right)} \label{eq:exponential},
\end{align}

\noindent
where $d_0 = 1.35~\text{\AA}$ represents the typical bonding distance within a DNA base. 

The constants $C_\chi$ are determined by fitting the method to experimental and ab initio results. QuantumDNA includes several parameterizations for these constants, with the default setting based on the latest MSF LCAO parameterization \cite{Mantela2021}.

To identify the HOMO and LUMO of each DNA base, we assume that all MOs below the Fermi level are occupied according to the Pauli exclusion principle, with each energy level accommodating two electrons due to spin degeneracy. Oxidation of a DNA base removes an electron from the HOMO, leaving a positively charged hole. Conversely, reduction adds an electron to the LUMO. Photoexcitation produces both an excited electron in the LUMO and a hole in the HOMO.

To estimate the CT parameters for both electrons and holes between neighboring DNA bases or base pairs, we compute the overlap of their MOs at the HOMO and LUMO levels. Fig.~\ref{fig:4_TB}(a) shows the overlapping MOs as shaded regions, representing molecular states that facilitate CT. The CT parameter between two adjacent molecules $A$ and $B$ is determined from this overlap as follows:

\begin{align} 
t_{AB} = \bra{\text{MO}_A} H_{\text{int}} \ket{\text{MO}_B},
\end{align}

\noindent
where $H_{\text{int}}$ is the molecular interaction Hamiltonian describing the coupling between molecules $A$ and $B$.

The following code demonstrates how to display the computed energies and CT parameters for two neighboring DNA bases, Adenine (A) and Thymine (T), derived from a given DNA geometry. In this example, the molecular geometry files are saved as \verb|A.xyz| and \verb|T.xyz| in the same directory. These files can be obtained from PubChem: Adenine (CID 190) and Thymine (CID 1135).

\begin{lstlisting}[language=python]
from qDNA import Base, load_xyz, Dimer

base_A = Base(*load_xyz("A"))
base_B = Base(*load_xyz("T"))
dimer = Dimer(base_A, base_B)

print("TB parameters")
print("-------------------------------")
print(f"E_HOMO_A: {base_A.E_HOMO}")
print(f"E_LUMO_A: {base_A.E_LUMO}")
print(f"E_HOMO_B: {base_B.E_HOMO}")
print(f"E_LUMO_B: {base_B.E_LUMO}")
print(f"t_HOMO: {dimer.t_HOMO}")
print(f"t_LUMO: {dimer.t_LUMO}")
\end{lstlisting}

Once the parameters for all the neighboring bases of the DNA segment under analysis have been evaluated, both for the HOMO and LUMO orbitals, they can be used as parameters in a coarser TB model (see Fig.~\ref{fig:4_TB}(b)) to model the dynamics of a charge along its double-strand.

\subsection{Tight-Binding (TB) Models}

The DNA molecule is a very complex system with many degrees of freedom, making a complete theoretical quantum description of it neither feasible nor practical. In order to reduce the complexity of such an analysis, the use of TB models, which have been developed to capture the most relevant features of a complex system while discarding less important ones, is well established. In DNA CT simulation, TB models have been used in many different studies, see Refs.~\cite{Chakraboty2007, Lambropoulos2019, Bittner2006,Herb2024,Rossini2024} and references therein.

TB models simplify the complex quantum mechanical problem at hand by introducing a coarse-grained approach that reduces each DNA base or base pair to a single degree of freedom, i.e., a site in the TB model. The couplings between neighboring sites are evaluated by considering the MO overlap between neighboring DNA bases, thus generating an effective model that captures the dynamics of a charge along different sites of the DNA strand, neglecting the charge dynamics within the atoms of each site. Since each nucleobase contains 13 to 16 atoms with multiple AOs, this approach reduces complexity by several orders of magnitude and thus significantly reduces the computational time needed to study larger DNA segments that are inaccessible by the ab initio methods mentioned above. TB models combine computational efficiency with physical accuracy, paving the way for the systematic analysis of large ensembles of DNA sequences.

\begin{figure*}[ht]
\centering
\includegraphics[width=1.\linewidth]{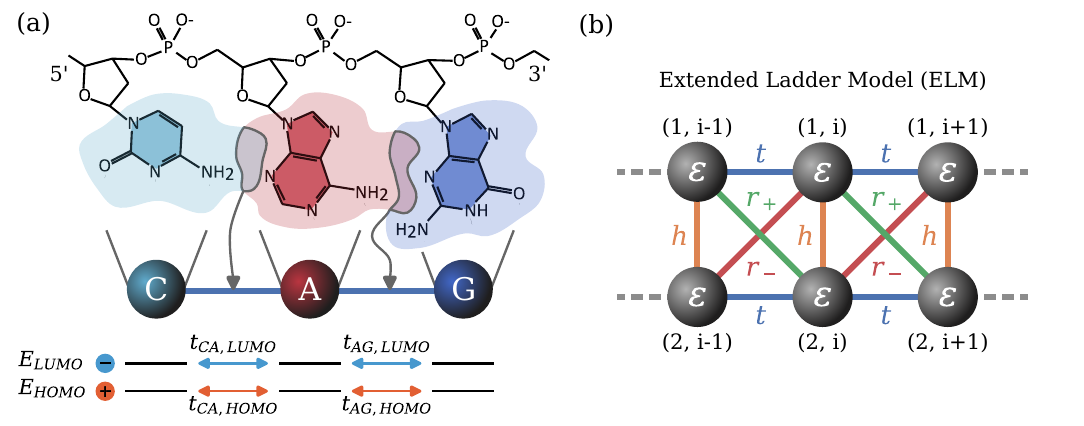}
\caption{\label{fig:4_TB} 
From the DNA molecule to TB modeling. 
\textbf{(a)} The chemical structure of the upper strand of the DNA sequence 5'-CAG-3', including the sugar-phosphate backbone, is shown. Each DNA nucleobase is simplified to its highest occupied molecular orbital (HOMO) and lowest unoccupied molecular orbital (LUMO). Overlap between these MOs, shown as shaded areas with gray edges, couples the HOMO and LUMO levels of neighboring bases, enabling CT.
\textbf{(b)} The extended ladder model (ELM) describes intrastrand transfer ($t$), vertical interstrand transfer within a base pair ($h$), and diagonal interstrand transfer ($r_{\pm}$) to account for the helical symmetry. The HOMO and LUMO energies are denoted by the on-site energy $\varepsilon$.}
\end{figure*}

The package includes several TB models for different levels of accuracy in DNA simulations, which are visualized in Fig.~\ref{fig:appendix_TB} and described below:

\begin{itemize}
\item The \emph{Wire Model (WM)} simplifies DNA by representing each base pair as a single TB site, providing a description at the base pair level. This model is useful for studying general DNA transport and conductance properties, but does not account for strand-specific dynamics. 

\item The \emph{Ladder Model (LM)} provides higher resolution by treating each DNA base as an individual TB site, thus capturing details at the single base level. It includes vertical hopping interactions within base pairs, but simplifies the DNA structure by representing it as a linear molecule, ignoring its helical symmetry. 

\item The \emph{Extended Ladder Model (ELM)} builds on the LM by incorporating the helical symmetry of DNA. This is achieved by including diagonal interstrand CT, resulting in a more accurate structural representation of DNA. As it is a popular TB model for DNA, QuantumDNA uses it by default, as shown in Fig.~\ref{fig:4_TB}(b). 

\item \emph{Fishbone models (FWM / FLM / FELM)} take complexity a step further by adding explicit modeling of the sugar-phosphate backbone, shown as brown rectangles above and below the strands in Fig.~\ref{fig:appendix_TB}. This improved representation captures the influence of the backbone on DNA dynamics, improving accuracy at the cost of increased computational complexity.
\end{itemize}

As shown in Fig.~\ref{fig:appendix_comp_time} in the appendix, the complexity and computational time required by each TB model increases, from top left to bottom right (in the legend). This is due to the different number of TB sites and number of CT parameters required to build each of those models.

QuantumDNA assigns a tuple $(i, j)$ to each TB site, where $i$ is the strand index and $j\in\{1, ..., N\}$ is the TB site index within the strand, given a strand length of $N$. These tuples form a set $\Lambda$, which represents the collection of the TB lattice sites. The local state $\ket{p_{i,j}}$ describes the presence of the particle $p$ (electron, hole or exciton) at the TB site indicated by the tuple $(i,j)$. Each TB Hamiltonian consists of two parts: a localization part containing the energies of the bases (HOMO and LUMO estimated energies) and a tunneling part containing the CT parameters (from the overlapping MOs of neighboring bases):

\begin{align}
    H^p_{\text{TB}} = H_{\text{loc}}^p + H_{\text{tun}}^p. 
\end{align}

The equations for the different TB models in Fig.~\ref{fig:appendix_TB} are explicitly given Eqs.~\eqref{eq:appendix_TB} in the appendix. 

\subsubsection{Coulomb and Exchange Interaction} 

When UV light photoexcites an electron from a lower-energy orbital to a higher-energy orbital in a DNA base, it leaves behind a positively charged hole that binds with the electron and forms a paired system often referred to as \emph{exciton}. Such state is described by the electron-hole configuration state $\ket{\text{e}_{\alpha_1} \text{h}_{\alpha_2}} $, where $\alpha_k = (i_k, j_k)$. The package includes two different types of two-body interactions, as described by Bittner \cite{Bittner2006, Bittner2007}, between the electron and the hole: the Coulomb interaction $J(r)$, representing the electron-hole attraction, and the exchange integrals $K(r)$, which account for spin-exchange coupling. Assuming that the Coulomb interaction decays algebraically and the exchange interaction decays exponentially with the distance $r$ between the charges, we implement these energy terms as \cite{Bittner2006, Bittner2007, Herb2024}: 

\begin{align}
    J(r) &= J_0 / (1+r/r_0) \\
    K(r) &=  K_0 \mathrm{e}^{-r/r_0}. 
\end{align}

\noindent 
where it is assumed the value $r_0=1.0\,\text{\AA}$ for the e-h separation within the same base \cite{Bittner2006,Bittner2007}. The distance among different bases is then evaluated considering that two adjacent DNA bases are separated by about $D=3.4\,\text{\AA}$, both intrastrand and interstrand. 

In addition to single CT, the exciton can be transferred along the DNA strand because of dipolar couplings between separated sites as described in \cite{Bittner2006}. Including the above introduced terms, the excited state Hamiltonian yields:

\begin{align}
    H_{\text{exc}} &= H_{\text{TB}}^e\otimes \mathbb{I}_h + \mathbb{I}_e\otimes H_{\text{TB}}^h + H_{\text{int}}.
\end{align}

As a default choice, the QuantumDNA package implements a cut-off in the evaluation of the Coulomb interaction to the nearest neighbouring bases. This is justified because the thermal energy at physiological temperatures of $300~\text{K}$ is about $0.3~\text{eV}$, which is equal to the Coulomb energy at the base-stacking distance $D$.

Normally, to account for such many-body interaction terms, it is necessary to explore an exponentially large Hilbert space with respect to the dimension of the system (number of DNA bases). However, the QuantumDNA package enforces the constraint that there is always either one exciton or no exciton in the system, allowing us to explore only a Hilbert space scaling quadratically in the length of the chain, i.e. $|\mathcal{H}|=|\Lambda|^2$.

The code below gives an example of how to retrieve the full excited state Hamiltonian of a GC/CG dimer once its parameters have been set:

\begin{lstlisting}[language=python]
from qDNA import DNA_Seq, TB_Ham

ham_kwargs = dict(coulomb_param = 1., exchange_param = 1.)
dna_seq = DNA_Seq('GC', 'ELM')
tb_ham = TB_Ham(dna_seq, **ham_kwargs)
tb_matrix = tb_ham.get_matrix()

print(f"TB Hamiltonian: {tb_matrix}")
\end{lstlisting}

\subsubsection{Isolated DNA Dynamics}

QuantumDNA provides useful tools for investigating the CT properties of isolated DNA chains, based on the evaluation of the dominant frequencies of charge oscillations along the chain. As the experimental study and manipulation of isolated DNA chains has become more relevant, as in the case of DNA origami \cite{Rothemund2006}, such theoretical tools can be extremely useful to analyze the intrinsic properties of DNA, free from the complex environmental interactions present in living cells, and provide insights into the timescales of charge dynamics and average charge distributions, as investigated by Lambropoulos et al. \cite{Lambropoulos2016a, Lambropoulos2016b}. 

In this context, the unitary dynamics of the system is governed by the TB Hamiltonian $H_{TB}$ via the von Neumann equation. For simplicity of notation, we use $\ket{\alpha}$ to represent both one-particle (1P) and two-particle (2P) states. The evolution of the system can be solved analytically and represented in terms of the eigenstates of the system $\ket{\Psi_i}=\sum_{\alpha\in\Lambda} c_{\alpha,i} \ket{\alpha}$ and the corresponding eigenenergies $E_i$. The time-dependent population $P_\alpha(t)$ of a state $\ket{\alpha}$ is then given by:
\begin{align}
   P_\alpha(t) &= \sum_{i=1}^N c_{i\alpha}^2 c_{i\beta}^2 + \sum_{i,j=1,\, i< j}^N 2\,c_{i\alpha} c_{i\beta} c_{j\alpha} c_{j\beta}\, \cos(\omega_{ij}t).
\end{align}
where $\omega_{ij} = (E_i-E_j)/\hbar$ represents the frequency associated with the energy difference between eigenstates $E_i$ and $E_j$. The first term represents the time-averaged population for non-degenerate eigenenergies, while the second term accounts for the oscillations between different eigenstates and their corresponding amplitudes.

A detailed knowledge of the expected dynamics of a charge along the DNA chain, both in terms of long-term behavior and fast scale oscillations, can be particularly useful in interpreting experimental measurements that are averaged over time, especially when ultrashort timescales are not directly resolvable.

The code below provides an example of how to perform the analysis described above for the GC/CG double-stranded segment:

\begin{lstlisting}[language=python]
from qDNA import DNA_Seq, TB_Ham

kwargs=dict(description='1P', particles=['hole'])
dna_seq = DNA_Seq('GC', 'ELM')
tb_ham = TB_Ham(dna_seq, **kwargs)

# Amplitudes and frequencies
tb_ham.get_amplitudes('(0, 0)', '(0, 0)')
tb_ham.get_frequencies('(0, 0)', '(0, 0)')

# Average populations
tb_ham.get_average_pop('(0, 0)', '(0, 0)')
\end{lstlisting}

\subsection{DNA Environment}

CT within DNA in the cell under physiological conditions is influenced by the cellular environment. Interactions with the solvent, especially water molecules, and counterions can induce fluctuations in the energies of the nucleobases, alter the conformation of the MOs of each DNA base, and inhibit the quantum nature of charge transport due to decoherence and relaxation phenomena. Thus, a thorough analysis of the complex interplay between the DNA structure, its dynamical fluctuations and the surrounding environment is crucial for the development of realistic models of DNA electronic properties and their implications in living biological processes.  

Several quantum physical models have been proposed to account for this effect \cite{Kubar2008, Gutierrez2009, Gutierrez2010, Rossini2024}. Often the environment is modeled as an infinite set of bosonic harmonic oscillators interacting with electronic transitions (excitons) within the DNA molecule \cite{Caldeira1983}. To describe the time evolution of a charge on the DNA strand without explicitly describing all the degrees of freedom of the bath, the influence of the bath on the particle has to be considered indirectly by solving an appropriate quantum master equation. This approach allows the effects of the environment on the charge dynamics along the DNA to be effectively represented, enabling the study of CT and excited state dynamics in biologically relevant systems.

To implement the above description in our modeling, the QuantumDNA package implements a set of Lindblad operators. Lindblad models are an efficient choice for simulating long DNA sequences due to their computational efficiency, ease of implementation, and analytical solvability for some specific systems. The Lindblad form of a quantum master equation for the time evolution of the density matrix $\rho(t)$ describing the charge dynamics is defined as:

\begin{align}
    \frac{\mathrm{d}}{\mathrm{d}t}\,\rho(t) = -\frac{\mathrm{i}}{\hbar}[H_{TB},\rho] + \sum_{k} L_{k}\rho L_{k}^{\dagger} - \frac{1}{2} \left(L_{k}^{\dagger}L_{k}\rho + \rho L_{k}^{\dagger}L_{k} \right)
\end{align}

\noindent
where the first term describes the coherent unitary evolution of the TB Hamiltonian of the isolated DNA described in the previous subsection, and the Lindblad operators $L_k$ capture the dissipative dynamics induced by environmental interactions. The relationship between the number of Lindblad operators and the number of TB sites (DNA sequence length) for each Lindblad model in the 2P description is illustrated in Fig.~\ref{fig:appendix_scaling} in the appendix.

The package implements four distinct methods for incorporating environmental effects, summarized in Tab.~\ref{tab:lindblad_ops}: (1) local dephasing, (2) global dephasing, (3) local thermalization, and (4) global thermalization. Dephasing models describe the loss of quantum coherence (decoherence) due to entanglement with the surrounding environment. However, unlike thermalization models, they do not drive the system toward a canonical thermal equilibrium state. In local models, each site interacts with an independent environment (bath), whereas global models assume a shared environment for all sites. For more details and benchmarking against more advanced beyond-Lindblad approaches, we refer to \cite{Abbott2020}.

\renewcommand{\arraystretch}{1.2} 
\begin{table*}[ht]
    \centering
    \captionsetup{justification=centering}
    \caption{Overview of Lindblad models implemented in QuantumDNA for the 1P description. The eigenstates of the DNA TB Hamiltonian are denoted as $\ket{\Psi_i}$ and the corresponding eigenfrequencies follow from the system's eigenenergies as $\omega_i = E_i/\hbar$. In the table, we use the conventions $N=|\Lambda|$, $w_{ij}= w_i-w_j$ and $c_{\alpha,i} = \braket{p_\alpha|\Psi_i}$. }
    \label{tab:lindblad_ops}
\begin{tabular}{@{}p{5cm}p{5cm}p{5cm}@{}}
    \toprule
    \textbf{Model} & \textbf{Operator} & \textbf{Conditions} \\
    \midrule
    Local dephasing \cite{Worster2019, Abbott2020, Rossini2024} & $\sqrt{\Gamma_{\alpha}^{\text{loc}}} \, \ket{p_{\alpha}} \bra{p_{\alpha}}$ & $\alpha \in \Lambda$ \\
    Global dephasing \cite{Rossini2024} & $\sqrt{\Gamma_{i}^{\text{glob}}} \, \ket{\Psi_{i}} \bra{\Psi_{i}}$ & $i = 1, \dots, N$ \\
    Local thermalizing \cite{Mohseni2008, Abbott2020} & $\sum\limits_{\substack{\omega=\omega_{ij}}} \sqrt{k(\omega)} \, c_{\alpha,j}^{*} c_{\alpha,i} \ket{\Psi_{j}} \bra{\Psi_{i}}$ & $\alpha \in \Lambda$ and $\omega \in \text{uniq}\{\omega_{ij}\}$ \\
    Global thermalizing \cite{Worster2019, Abbott2020} & $\sqrt{k(\omega_{ij})} \, \ket{\Psi_{j}} \bra{\Psi_{i}}$ & $i, j = 1, \dots, N$ and $i \neq j$ \\
    \bottomrule
\end{tabular}
\end{table*}

\subsection{Excitation Relaxation Channel}

The DNA molecule is inherently photostable due to the rapid decay of its electronic excitations. To model this relaxation process, we adopt a minimalist approach inspired by pump-probe experiments as in \cite{Crespo-Hernandez2004} and theoretically introduced in \cite{Herb2024}. To account for the possibility that the DNA molecule relaxes to its ground state after excitation (creation of an e-h pair), the Hilbert space of the system is extended to include the ground state of DNA, labeled $\ket{0}$. In addition, intrinsic loss mechanisms are incorporated using Lindblad operators that describe the possibility of exciton recombination occurring at each site of the TB model with a tunable decay rate $\gamma_\alpha$:

\begin{align}  	L_{\alpha}=\sqrt{\gamma_{\alpha}}\ket{0}\bra{e_{\alpha}h_{\alpha}}.\, 
    \label{eq:recombination}
\end{align}

We identify the exciton lifetime as the time taken for the excitation to decay by $\epsilon=1/\text{e}$ of its initial value, following \cite{Herb2024}. Specifically, we define the exciton lifetime $T$ as the shortest time for which the population satisfies the condition:

\begin{align}
    P_0(T) = \bra{0}\rho(T)\ket{0} \geq 1 - \epsilon.
\end{align}

In addition to describing the lifetime of the e-h pair, it is possible to quantify the separation between the electron and the hole (charge separation) occurring due to the dynamics of each particle using the charge separation operator \(\hat{d}\), defined in our package as:

\begin{align}
   \bra{e_{\alpha_1} h_{\alpha_2}}  \hat{d} \ket{e_{\alpha_1} h_{\alpha_2}} = D \cdot \left(|i_1-i_2|+|j_1-j_2|\right),
\end{align}
where $\alpha_1=(i_1,j_1)$, $\alpha_2=(i_2,j_2)$, and $D = 3.4~\text{\AA}$ corresponds to the average distance between adjacent DNA base pairs. The time-averaged charge separation, $\bar{d}$, provides an estimate of the amount of charge separation over the exciton lifetime, $T$:
\begin{align}
    \bar{d} = \frac{1}{T}\int_0^{T} \text{d}t\, \text{Tr}\{\rho(t)\, \hat{d}\}.
\end{align}

The amount of separation between the electron and the hole can affect the electronic properties of DNA, as the two charges generate an electric dipole characterized by an effective dipole moment. This dipole moment can influence interactions with other molecules, such as proteins, which may be critical to the biological function of DNA.

\section{Application and Benchmarking Examples} \label{sec:features}

The development of QuantumDNA was driven by the observation that many research studies employ similar methodologies, such as TB models, to explore the quantum physical properties of DNA. This motivated the need for a unified platform to standardize and streamline these approaches. QuantumDNA therefore integrates methods from different research groups and has been benchmarked against published results.

In this section, we will use several examples to demonstrate the variety of different analyses that can be performed with the QuantumDNA package using just a few lines of code and only built-in functions. In particular, we will show how this package succeeds in converging the efforts of different research groups already working on this topic, but often pursuing independent research strategies. We then go on to show how this package paves the way for novel investigative strategies that can be integrated with those already existing in the field, thus setting the stage for more biologically oriented investigations.

\subsection{Reproduction of Ultrafast Excitonic Dynamics along the DNA}

\begin{figure*}[ht]
\centering
\includegraphics[width=1.\linewidth]{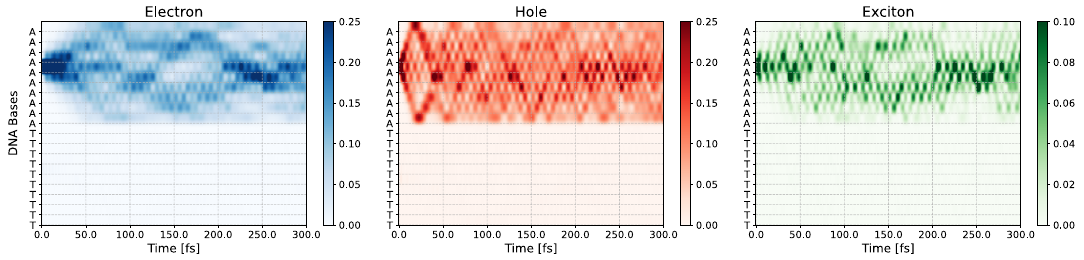}
\caption{\label{fig:6_reproduction_Bittner}Electron, hole and excitonic dynamics over a DNA modeled TB lattice for 300 fs. In the plots, an electron-hole pair is first generated over the fifth adenine of an ELM consisting of 10 adenine bases and their opposite thymine bases, following the modeling and parameterization introduced in \cite{Bittner2006, Mehrez2005}. Here we add the dynamical information about the independent dynamics of the electron (blue) and the hole (red), which together produce the excitonic pattern in green.}
\end{figure*}

In 2006 Eric Bittner presented one of the first theoretical investigations of the ultrafast dynamics of an exciton along the DNA chain \cite{Bittner2006,Bittner2007}, using a TB model parameterized with ab initio calculations from Mehrez and Anantram \cite{Mehrez2005}, also available in this package (source identifier \texttt{Bittner2007}). In his model, in contrast to several investigations carried out later by the community, he already includes a strategy to take into account the effects of the Coulomb and spin-exchange interactions that occur between the electron and the hole when both are present on the chain.

With the QuantumDNA package it is possible to reproduce the results developed by such modeling and to study the dynamics of electrons and holes along the chain under the influence of the same effects. In addition to the excitonic dynamics, it is possible to visualize independently the dynamics of the individual charges.

The example in Fig.~\ref{fig:6_reproduction_Bittner} shows the results produced by the QuantumDNA package for a similar scenario considered in \cite{Bittner2006}, with an exciton initially placed over the fifth adenine of a double-stranded model consisting of 10 adenines (A) and 10 opposite thymines (T). The plot shows the dynamics for the electron, the hole, and the exciton. This plot can be easily obtained by running the code below: 

\begin{lstlisting}[language=Python]
from qDNA import get_me_solver, plot_pops_heatmap

upper_strand = 'AAAAAAAAAA'
tb_model_name = 'ELM'
kwargs = dict(relaxation=False, 
    source='Bittner2007', unit='eV', 
    coulomb_interaction=2.5, 
    exchange_interaction=1, 
    init_e_state='(0, 4)',
    init_h_state='(0, 4)', 
    t_end=300, t_unit='fs', t_steps=300)

me_solver = get_me_solver(upper_strand, tb_model_name, **kwargs)

fig, ax = plot_pops_heatmap(me_solver,
    heatmap_type='seaborn', 
    vmax_list=[0.25, 0.25, 0.1])
\end{lstlisting}

The extent of these results can be generalized by modifying the input DNA sequence and system parametrization as explained in Section~\ref{sec:package_structure}. For example, by allowing the exciton to annihilate and thus escape the DNA strand, it is possible to model experimental results from Kohler et al. \cite{Crespo-Hernandez2005} to investigate the underlying mechanisms for the long-lived excitations measured in their pump-probe experiments on different DNA segments. This approach is described in more detail in \cite{Herb2024}.

\subsection{Modeling the Transition between Superexchange and Classical Hopping CT Regimes}

\begin{figure}[ht]
\centering
\includegraphics[width=1.\linewidth]{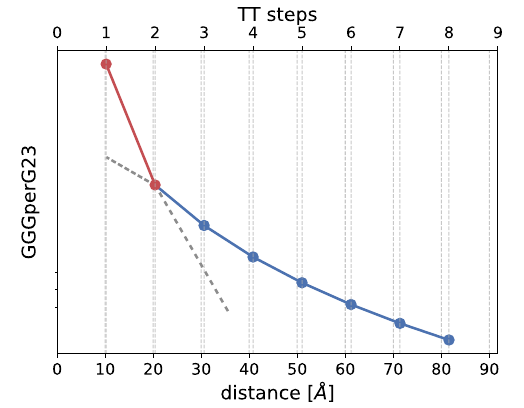}
\caption{\label{fig:6_reproduction_Simserides} Semi-logarithmic plot of the time-averaged hole population ratios at the donor (G23) and acceptor (GGG) sites. The sequences analyzed follow the format '[tail] G23 - bridge - GGG [tail]' with variable bridge lengths composed of TTGTT repeats, inspired by experiments from Giese et al. \cite{Giese1999, Giese2001}. The plot qualitatively demonstrates a crossover from the superexchange mechanism to the thermal hopping regime. Originally presented in \cite{Simserides2014}, this result was reproduced using QuantumDNA, with the corresponding code provided in main text.}
\end{figure}

Another example of the breadth of analysis provided by QuantumDNA is shown in Fig.~\ref{fig:6_reproduction_Simserides}. Pioneering experiments by Giese et al. \cite{Giese1999, Giese2001} observed a transition from quantum mechanical tunneling to classical thermal hopping in hole transport under varying conditions. They studied the sequence dependence of the transfer rates of a hole from its initial position on a donor base (G23) to an acceptor complex (triplet GGG) across a bridge of varying length, and deduced from the distribution of results that different CT mechanisms must take place depending on the bridge length. This phenomenon was later modeled theoretically by Simserides et al. \cite{Simserides2014}, whose simulations qualitatively reproduced the crossover found in Giese's experiments. QuantumDNA incorporates this modeling strategy, allowing these theoretical results to be reproduced using the code below, and extended to more complex or biologically relevant systems.

\begin{listing}
\begin{lstlisting}[language=Python]
import matplotlib.pyplot as plt
from qDNA import TB_Ham, DNA_Seq

tb_model_name = 'WM'
kwargs = dict(description='1P', particles=['hole'], unit='rad/fs',
              relaxation=False, source='Simserides2014')

front_tail = 'ACGCACGTCGCATAATATTAC'
back_tail = 'TATTATATTACGC'
GGG_per_G = []
for num_steps in range(1, 10):
    bridge = 'TT' + 'GTT' * (num_steps-1)
    upper_strand = front_tail+'G'+bridge+'GGG'+back_tail
    dna_seq = DNA_Seq(upper_strand, tb_model_name)
    tb_ham = TB_Ham(dna_seq, **kwargs)
    
    donor_site = '(0, 21)' # guanine left from the bridge    
    acceptor_sites = [
        f'(0, {21+len(bridge)+1})', 
        f'(0, {21+len(bridge)+2})',  
        f'(0, {21+len(bridge)+3})']

    donor_avg_pop = tb_ham.get_average_pop(donor_site, donor_site)['hole']
    acceptor_avg_pop = 0
    for acceptor_site in acceptor_sites:
        acceptor_avg_pop += tb_ham.get_average_pop(donor_site, acceptor_site)['hole']

    GGG_per_G_ratio = acceptor_avg_pop / donor_avg_pop 
    GGG_per_G.append(GGG_per_G_ratio)

plt.plot(list(range(1,10)), GGG_per_G)
\end{lstlisting}
\end{listing}

\begin{figure*}[ht]
\centering
\includegraphics[width=1.\linewidth]{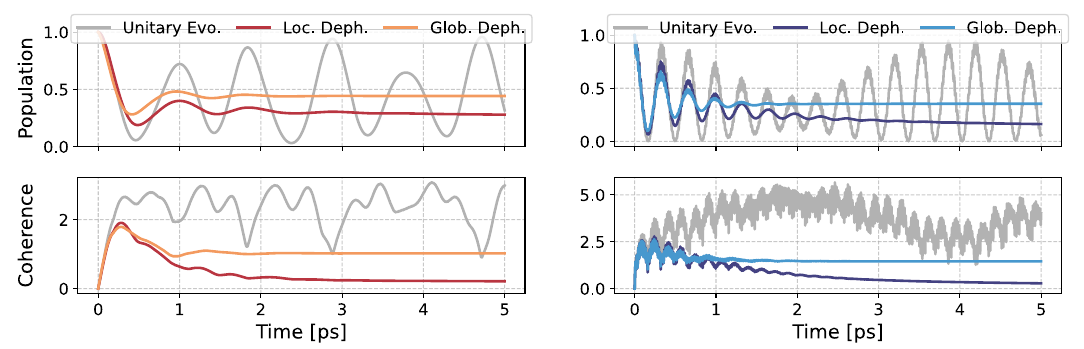}
\caption{\label{fig:6_isolated_open} Population and coherence dynamics of a hole (left) and an electron (right) in a DNA TB model. The plots illustrate the particle dynamics over a Ladder Model (LM) representation of a GCACG DNA strand under three different conditions: isolated system (unitary evolution), local dephasing, and global dephasing.
In the upper panels, we show the population dynamics of a particle initially localized on the top-left guanine base. For clarity, only the dynamics of this base are displayed, while the behavior of other bases is omitted (see \cite{Rossini2024} for a comprehensive discussion).
The lower panels depict the coherence dynamics of the entire system, highlighting the decay of quantum coherence due to environmental interactions. The differences between local and global dephasing models become evident, illustrating how each approach affects the coherence decay over time.}
\end{figure*}

\subsection{Quantum Charge Dynamics over an Isolated or Open DNA Chain}

The quantum dynamics of a charge on a DNA strand can be strongly influenced by the presence of an environment. In particular, the interaction of a quantum system with its environment is known to lead to a loss of coherence of the system, i.e. the tendency of the system to switch from quantum to classical behavior. Therefore, in order to understand the possible impact of quantum effects on biologically relevant systems, it is of utmost importance to carefully study the effects of the environment on the dynamics of a quantum system. Such an analysis, introduced by the work of Rossini et al.\cite{Rossini2024}, can be easily performed using the QuantumDNA package, as introduced in section~\ref{sec:package_structure} and illustrated in Fig.~\ref{fig:6_isolated_open}. The system studied in the example is a double-stranded GCACG segment in which an electron or hole is initially placed on the LUMO or HOMO of the upper left guanine base. The top plots show the population of particles on this guanine over time for three different scenarios: (1) unitary system -- the DNA is isolated; (2) local coupling with the environment -- each site of the DNA strand interacts independently with the surrounding environment; (3) global coupling with the environment -- the environment interacts with global modes (or eigenstates) of the DNA system.

The dynamics of the particles along the DNA chain varies depending on the scenario. Particles on isolated DNA oscillate continuously, while particles on interacting DNA gradually damp as they interact with the environment, reflecting the loss of quantum properties. The lower plots in Fig.~\ref{fig:6_isolated_open} show a measure of the coherence, i.e. "quantumness", of the system, the sum of the absolute values of the off-diagonal terms of the particle density matrix. We emphasize that there are many ways to investigate how quantum a state is \cite{Baumgratz2014}, and that the QuantumDNA package is structured so that any preferred measure can be implemented to replace the built-in ones. The coherence measure of the unitary dynamics oscillates, maintaining the same recurring peak level, showing how the quantumness of the system is preserved over time. The plots for the interacting scenarios instead show a gradual loss of coherence to a final plateau, corresponding to the final equilibrium of the system. Further details can be found in \cite{Rossini2024}, while the code for the plot in Fig.~\ref{fig:6_isolated_open} is given below:

\begin{lstlisting}[language=Python]
import matplotlib.pyplot as plt
from qDNA import get_me_solver, plot_pop, plot_coh

upper_strand = 'GCACG'
tb_model_name = 'LM'
tb_site = '(0, 0)'

# change particle to ['electron'], loc_deph_rate and glob_deph_rate
kwargs = dict(description='1P', t_end=5, 
        particles=['hole'],
        loc_deph_rate=2,
        glob_deph_rate=0)
me_solver = get_me_solver(upper_strand, tb_model_name, **kwargs)

fig, ax = plt.subplots(2, 1)
plot_pop(ax[0], tb_site, me_solver)
plot_coh(ax[1], me_solver)
\end{lstlisting}

Modeling the nature of quantum CT dynamics on DNA in the presence of an environment is crucial for any biologically relevant investigation, and finding the best modeling strategy for this task is still an open question, which the QuantumDNA package allows to address efficiently.

\subsection{Effects of Mutations of the TERT DNA Sequence on Charge Dynamics}

\begin{figure}[H]
\centering
\includegraphics[width=1.\linewidth]{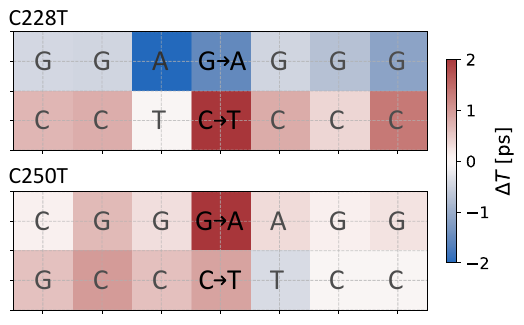}
\caption{\label{fig:6_TERT}Effect of TERT sequence mutations on the average exciton lifetime across the DNA strand. The heatmaps illustrate how exciton lifetimes change when the C228T and C250T mutations occur in the TERT promoter region, as described in \cite{Huang2013, Horn2013, Min2016}. Each site on the heatmap represents the variation in lifetime when the exciton is initially localized at that position. The C228T mutation decreases exciton lifetimes on the upper strand while promoting longer-lived excitations on the lower strand. In contrast, the C250T mutation generally leads to longer-lived excitations, which is typically associated with reduced system stability. These preliminary findings suggest potential implications for medical and epigenetic research, requiring further investigation to draw definitive conclusions.}
\end{figure}

Finally, we present an example of novel investigations that can be carried out using the QuantumDNA package. Here, we investigate a specific sequence within the human genome, the so-called hTERT sequence, for \emph{human Telomerase Reverse Transcriptase}. In 2013 two studies \cite{Huang2013,Horn2013} were published describing how two specific non-coding mutations of hTERT, located at the 1.295.228$^{th}$ and 1.295.250$^{th}$ base pair of the fifth chromosome, were recurrent in the appearance of melanomas. In both cases, a GC base pair mutates to a AT base pair \cite{Min2016}. They are usually referred to as C228T and C250T hTERT mutations.

The QuantumDNA package allows us to investigate the effect of such mutations on the quantum CT properties of DNA segments containing them. In particular, we investigate their effect on the expected lifetime of excitons initially excited by external UV radiation, as melanoma is a well-known skin cancer. For both mutations, we modeled a double-stranded ELM-TB model containing three base pairs before and after the mutating one, for a total of 14 DNA bases. We then evaluated the average lifetime of an exciton initially located at each one of the available sites, in both the natural and mutant cases. Finally, in Fig.~\ref{fig:6_TERT} we visualize the differences between the lifetimes evaluated in the natural and mutated cases for each site of the double strand using a heatmap. It appears clear how the mutations affect the lifetime of excitons along the DNA chain, with particularly strong effects for excitations originating at or near the site of the mutation.

The results presented at this stage are not intended to validate an effective quantum influence on the mechanisms linking hTERT mutations to melanoma, but rather to provide an example (currently under investigation) of how the QuantumDNA package can drive research on quantum CT transfer along DNA towards more impactful avenues for the fields of epigenetics and medicine.

\section{Conclusion} \label{sec:conclusion}

In this paper, we present QuantumDNA, an open-source Python package for the numerical simulation and analysis of open quantum charge dynamics in DNA, which aims to bridge the gap between quantum physics, biology and medicine. This framework provides a versatile computational toolkit that integrates LCAO and TB models with quantum master equations, enabling researchers to explore charge transport, excitonic behavior and environmental effects in DNA. Through its modular architecture and easy-to-use GUI, QuantumDNA makes quantum simulations of DNA accessible to researchers from different disciplines, fostering interdisciplinary collaboration across multiple fields.

We have outlined the theoretical foundation of QuantumDNA, detailing its implementation of quantum CT models and its ability to reproduce known experimental and theoretical results. Furthermore, we have demonstrated its applicability by presenting benchmark simulations and biological case studies, including the implementation of known models from the literature and the investigation of possible effects of sequence mutations on exciton lifetimes. These examples illustrate the potential of QuantumDNA to provide new insights into genetic and epigenetic phenomena through quantum physical analysis. 

While this first release lays a solid foundation, future developments will focus on extending the capabilities of the package, including the incorporation of Green's function methods for improved transport calculations and more advanced LCAO parameterizations that explicitly include backbone-mediated CT, an area of growing interest in the community \cite{Zhuravel2020}. In addition, further refinement of environmental interactions will increase the biological relevance of QuantumDNA simulations. 

As an evolving open-source platform, QuantumDNA encourages community contributions to refine and extend its functionalities. By bridging quantum physics and computational biology, QuantumDNA paves the way for novel investigations of the electronic properties of DNA, offering new perspectives on its role in DNA repair, gene expression and regulation, and potential medical applications.

\section*{Acknowledgements}

The authors would like to thank Reiner Siebert and Ole Ammerpohl for fruitful discussions, especially regarding the medical and epigenetic applications of this package. We thank Paul Raschke for his contributions to the technical development of the TERT results. 

This work has been supported by the Center for Integrated Quantum Science and Technology (IQST) and the BMBF through the QCOMP project in Cluster4Future QSens.

\section*{Data Availability}

The source code of the package is freely available under the BSD 3-Clause License and can be accessed on GitHub at \href{https://github.com/dehe1011/QuantumDNA}{github.com/dehe1011/QuantumDNA}. The documentation webpage including the API and a user guide with tutorial Jupyter notebooks is available at \href{https://quantumdna.readthedocs.io/en/latest/}{quantumdna.readthedocs.io/en/latest/}. Tutorial Notebooks are also available at \href{https://github.com/dehe1011/QuantumDNA-notebooks}{github.com/dehe1011/QuantumDNA-notebooks}. Contributions to the code are welcome, especially for integrating interfaces with other tools.



\bibliographystyle{elsarticle-num}
\bibliography{biblio.bib}


\newpage
\onecolumn
\appendix
\section{Appendix} \label{sec:appendix}

\setcounter{table}{0}  
\renewcommand{\thetable}{A\arabic{table}} 
\setcounter{figure}{0} 
\renewcommand{\thefigure}{A\arabic{figure}}

\renewcommand{\arraystretch}{1.2}
\vfill
\begin{table}[H]
    \centering
    \captionsetup{justification=centering}
    \caption{Description of selected classes and functions. For a more detailed description of the parameters, attributes and methods we refer to the \href{quantumdna.readthedocs.io/en/latest/}{QuantumDNA documentation webpage}. }
    \label{tab:classes}
    \begin{tabular}{@{}p{3cm}p{11cm}@{}}
        \toprule
        \textbf{Class/ Function} & \textbf{Description}  \\
        \midrule
        \verb|Base| & This class calculates the HOMO and LUMO and their energies from the geometry of a DNA base using Slater-Koster expressions for atomic orbital overlap. \\
        
        \verb|BasePair| & This class combines DNA bases that form a base pair to calculate the HOMO and LUMO and their energies for a base pair. \\
        
        \verb|Dimer| & This class combines two molecules (DNA bases or DNA base pairs) and calculates the transfer integral for the electron (LUMO) and hole (HOMO) and mediated by the dimer Hamiltonian evaluates the overlap of the molecular orbitals. \\
        
        \verb|DNA_Seq| & This class contains information about the DNA sequence tailored to the selected TB model by assigning DNA bases to TB sites, thus bridging the classes \verb|TB_Model| and \verb|TB_Ham|. \\
        
        \verb|TB_Model| & This class automatically retrieves the appropriate model
        properties and configurations from the resources of the package, given the TB model identifier and the dimensions of the system (tuple with the number of strands and sites per strand). \\

        \verb|TB_Ham| & This class builds a TB Hamiltonian matrix for the DNA molecule using the TB model, TB sites and TB parameters for selected particles (electron, hole or exciton). Optionally, the DNA ground state, the electron-hole Coulomb interaction and the exchange interaction can be added. The class includes the ability to perform Fourier analysis and calculate time-averaged populations and charge transfer rates for non-dissipative unitary dynamics. \\
        
        \verb|Lindblad_Diss| & This class is designed for the construction of Lindblad operators
        for master equations to describe various dissipative processes such as intrinsic relaxation to the ground state, dephasing and thermalisation due to the thermal DNA environment. It also includes a set of observables covering populations (electron, hole, exciton and ground state) and coherences. \\
        
        \verb|ME_Solver| & This class combines the DNA TB model and its environment into a quantum master equation as shown in Fig.~\ref{fig:2_qDNA_structure}. This differential equation is solved by QuTiP's efficient master equation solvers \cite{qutip1, qutip2} for a selected initial charge state. \\

        \bottomrule
    \end{tabular}
\end{table}
\vfill\clearpage


\begin{figure}[p]
\centering
\includegraphics[width=1.\linewidth]{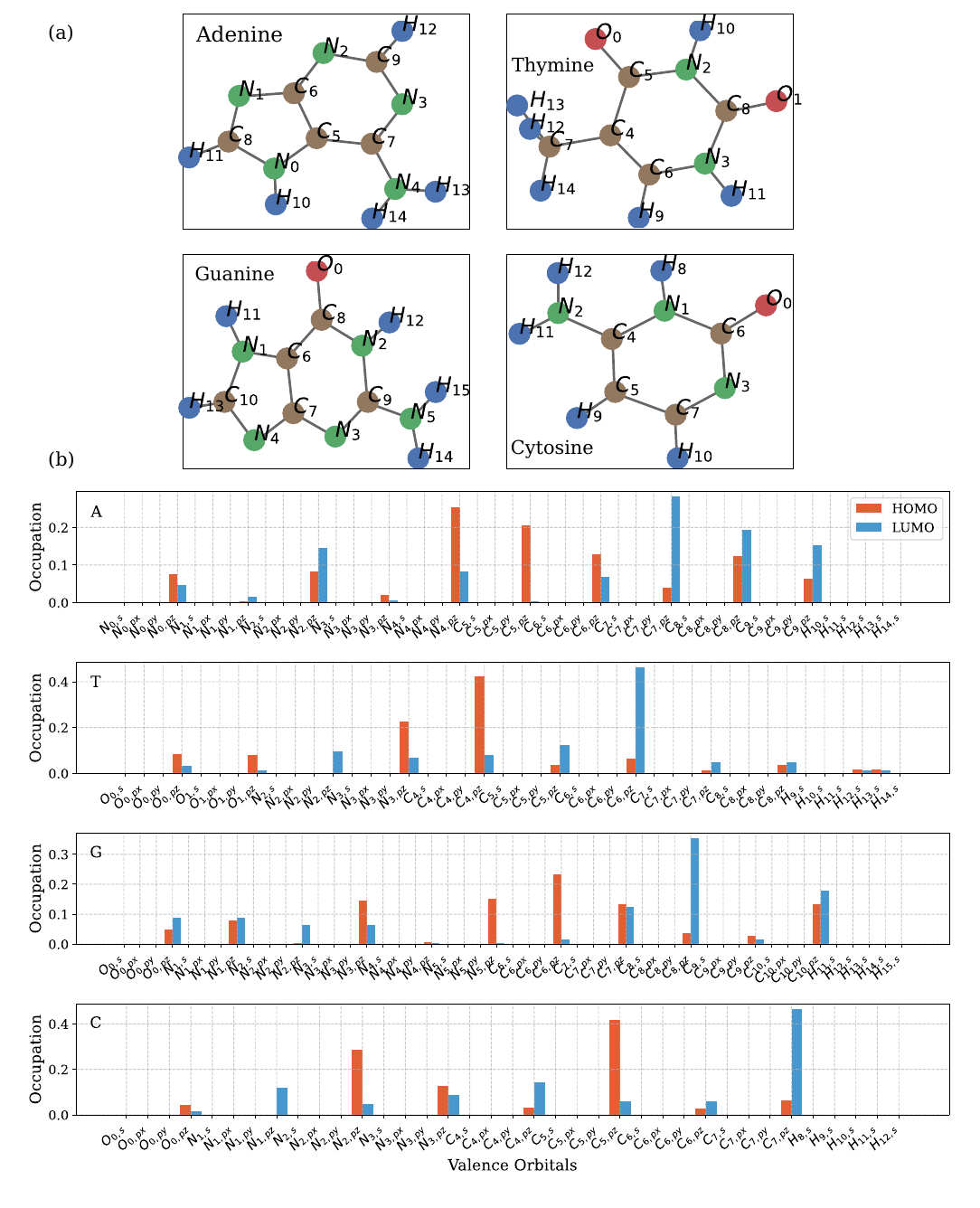}
\caption{\label{fig:appendix_LCAO} 
\textbf{(a)}  Visualization of the geometries of the DNA nucleobases obtained from \texttt{PubChem}: Adenine (CID 190), Thymine (CID 1135), Guanine (CID 135398634), and Cytosine (CID 597).
\textbf{(b)} Distribution of the HOMO and LUMO across valence orbitals based on the LCAO parametrization from \cite{Mantela2021}. The molecular orbitals predominantly occupy the pz orbitals, consistent with the established understanding that DNA charge transfer primarily occurs along the pi bonds. }
\end{figure}
\clearpage


\vfill
\begin{figure}[H]
\centering
\includegraphics[width=1.\linewidth]{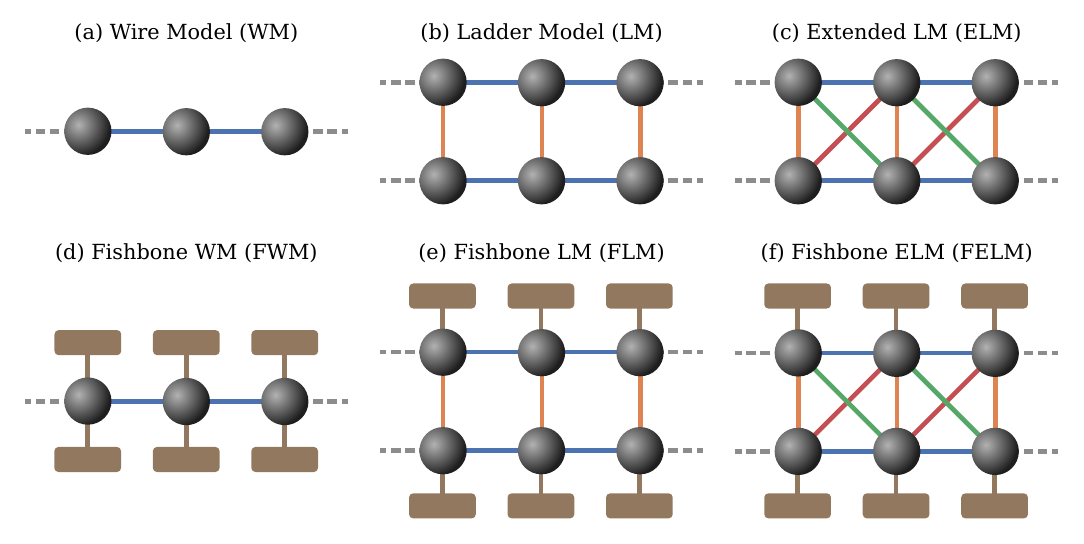}
\caption{\label{fig:appendix_TB}
Overview over different TB models for varying degrees of accuracy in DNA simulations. 
\textbf{(a)} Wire Model (WM): Each DNA base pair is represented as a single TB site, providing a base-pair-level description. This model is primarily used to study DNA transport and conductance but cannot resolve strand-specific dynamics.
\textbf{(b)} Ladder Model (LM): Each DNA base is treated as a separate TB site, allowing for single-base-level resolution. Vertical hopping within base pairs is included, but the model represents DNA as a linear molecule, ignoring its helical symmetry.
\textbf{(c)} Extended Ladder Model (ELM): This model incorporates the helical symmetry of DNA by including diagonal interstrand charge transfer, improving the representation of DNA structure.
\textbf{(d-f)} Fishbone Models: These models extend the previous approaches by explicitly including the sugar-phosphate backbone, depicted as brown rectangles above and below the strands. This more detailed representation provides greater accuracy but comes at the expense of increased computational complexity. }
\end{figure}

\begin{subequations}
\label{eq:appendix_TB}
\begin{align}
    H_{\text{WM}}^p &= H_{\text{loc}}^p + \sum_{j=1}^N t_{1,j}\, \ket{p_{1,j}} \bra{p_{1,j+1}} + \text{H.c.} \\
    H_{\text{LM}}^p &= H_{\text{loc}}^p + \sum_{j=1}^N \left( t_{1,j}\, \ket{p_{1,j}} \bra{p_{1,j+1}} + t_{2,j}\, \ket{p_{2,j}} \bra{p_{2,j+1}} + h_{1,j}\, \ket{p_{1,j}} \bra{p_{2,j}} \right) + \text{H.c.} \\
    H_{\text{ELM}}^p &= H_{\text{LM}}^p + \sum_{j=1}^N \left( r^+_{1,j}\, \ket{p_{1,j}} \bra{p_{2,j+1}} + r^-_{1,j}\, \ket{p_{1,j}} \bra{p_{2,j-1}} \right) + \text{H.c.} \\
    H_{\text{FWM}}^p &= H_{\text{WM}}^p + \sum_{j=1}^N \left( h_{0,j}\, \ket{p_{0,j}} \bra{p_{1,j}} + h_{1,j}\, \ket{p_{1,j}} \bra{p_{2,j}} \right) + \text{H.c.} \\
    H_{\text{FLM}}^p &= H_{\text{LM}}^p + \sum_{j=1}^N \left( h_{0,j}\, \ket{p_{0,j}} \bra{p_{1,j}} + h_{2,j}\, \ket{p_{2,j}} \bra{p_{3,j}} \right) + \text{H.c.} \\
    H_{\text{FELM}}^p &= H_{\text{ELM}}^p + \sum_{j=1}^N \left( h_{0,j}\, \ket{p_{0,j}} \bra{p_{1,j}} + h_{2,j}\, \ket{p_{2,j}} \bra{p_{3,j}} \right) + \text{H.c.}.
\end{align}
\end{subequations}

In the above expressions, TB parameters that describe transfer to TB sites not included in the set $\Lambda$ are set to zero. Specifically, we have $t_{i,j}=h_{i,j}=r^+_{i,j}=r^-_{i,j}=0\, \text{if}\, (i,j)\notin\Lambda$. 
\vfill\clearpage


\vfill
\begin{figure}[H]
\centering
\includegraphics[width=0.7\linewidth]{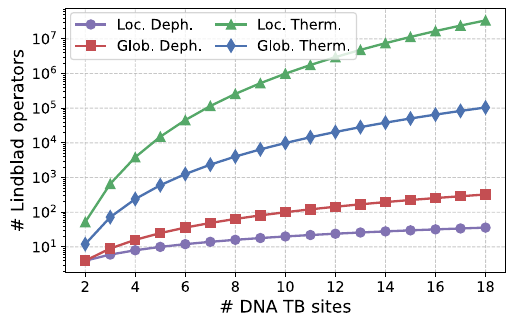}
\caption{\label{fig:appendix_scaling} Semi-logarithmic plot of the scaling of the number of Lindblad operators with the number of TB sites (DNA sequence length) for different Lindblad models. The scaling is shown for local dephasing (purple  circles, $2N$), global dephasing (red squares, $N^2$), local thermalizing (green triangles, $\leq N^2(N^4-N^2+1)$), and global thermalizing (blue diamonds, $N^4-N^2$), where $N = |\Lambda|$ denotes the number of TB sites.}
\end{figure}
\vfill
\begin{figure}[H]
\centering
\includegraphics[width=0.7\linewidth]{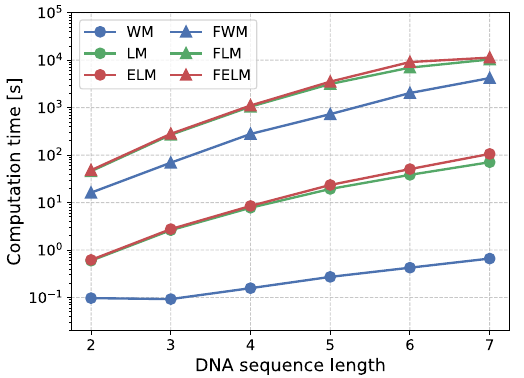}
\caption{\label{fig:appendix_comp_time} Computation time for the calculation of the exciton lifetime for different DNA sequence lengths for predefined TB models. We utilized a 12th Gen Intel Core i9-12900 processor for our computational simulations, which offers 16 cores (24 threads) and operates with 2.40 GHz Base Clock and 5.10 GHz Boost Clock. }
\end{figure}
\vfill\clearpage


\begin{table}[p]
    \centering
    \captionsetup{justification=centering}
    \caption{Overview of default values used by the package. The values can be modified in the \texttt{defaults.yaml} file located in the package's root directory. By default, the "MSF" parametrization is used to calculate a custom TB parameters source. Instead, when not used, the package defaults to the "Hawke2010" source.} 
    \label{tab:defaults}
    \begin{tabular}{@{}p{3cm}p{3cm}|p{3cm}p{3cm}@{}}
        \toprule
        \textbf{Variable} & \textbf{Default} & \textbf{Variable} & \textbf{Default} \\
        \midrule
        \verb|verbose| & False & \verb|loc_therm| & False \\
        \verb|parametrization| & "MSF" & \verb|glob_therm| & False \\
        \verb|source| & "Hawke2010" & \verb|deph_rate| & 7 \\
        \verb|description| & "2P" & \verb|cutoff_freq| & 20 \\
        \verb|particles| & ["electron", "hole", "exciton"] & \verb|reorg_energy| & 1 \\
        \verb|unit| & "rad/ps" & \verb|temperature| & 300 (K) \\
        \verb|coulomb_param| & 0 & \verb|spectral_density| & "debye" \\
        \verb|exchange_param| & 0 & \verb|exponent| & 1 \\
        \verb|relaxation| & True & \verb|t_steps| & 500 \\
        \verb|nn_cutoff| & True & \verb|t_end| & 3 \\
        \verb|loc_deph_rate| & 0 & \verb|t_unit| & "ps" \\
        \verb|glob_deph_rate| & 0 & \verb|init_e_state| & "(0, 0)" \\
        \verb|uniform_relaxation| & True & \verb|init_h_state| & "(0, 0)" \\
        \verb|relax_rate| & 0 & \verb|deloc_init_state| & False \\
        \verb|relax_rates| & \{"A": 0, "T": 0, "G": 0, "C": 0, "F": 0, "B": 0\} & \verb|solver_method| & "adams" \\
        \bottomrule
    \end{tabular}
\end{table}
\clearpage


\begin{table}[p]
    \centering
    \captionsetup{justification=centering}
    \caption{Sets of TB parameters predefined in QuantumDNA.}
    \label{tab:tb_params}
    \begin{tabular}{@{}p{3cm}p{3cm}p{9cm}@{}}
        \toprule
        \textbf{Source} & \textbf{TB Models} & \textbf{Method} \\
        \midrule
        Endres2004 \cite{Endres2002, Endres2004} & WM & TB parameters calculated by DFT and Slater-Koster model with a parametrization fitted to ab initio calculations. Limited to $p_z$ valence orbitals (Hückel-type LCAO) and 5'-GG-3', 5'-AA-3', 5'-AG-3', 5'-GA-3'.\\
        
        Bittner2007 \cite{Bittner2006, Bittner2007} & LM, ELM & Electron and hole TB parameters taken from Mehrez and Anantram \cite{Mehrez2005} calculated by DFT with B3LYP density functional and 6–31G basis set, limited to poly(G)-poly(C) and poly(A)-poly(T) DNA. Extended by dipole-dipole coupling terms calculated using a point-dipole approximation evaluated using configuration interaction singles (CIS) and Coulomb and Exchange interactions to account for exciton transfer.  \\
        
        Hawke2010 \cite{Hawke2010} & WM, LM, ELM & TB parameters calculated using a novel Slater-Koster parametrization fitted to available experimental data and ab inito simulations. Limited to $p_z$ valence orbitals (Hückel-type LCAO). \\
        
        Simserides2014 \cite{Simserides2014} & WM & Review and consolidation of existing sets of TB parameters into one consistent set for WM description on base-pair level. \\
        
        Mantela2021 \cite{Mantela2021} & WM & TB parameters calculated using a novel Slater-Koster parametrization including all valence orbitals to describe structural variability like deviations from the ideal planar geometry. Limited to description on base-pair level. \\
        
        Herb2024 & WM, LM, ELM & TB parameters calculated using the QuantumDNA package on \verb|PBD| files created with Biovia Discovery Studio using the parametrization from Mantela2021. Allows for a description on single-base level. \\

        1BNA & ELM & TB parameters tailored for simulations on subsequences within the 1BNA molecule (dodecamer 5'-CGCGAATTCGCG-3') obtained using the QuantumDNA package. The geometry found by x-ray diffraction was obtained from \verb|rcsb.org| with the identifier "1BNA" and modified using Biovia Discovery Studio. These parameters result from the approach shown in Fig.~\ref{fig:5_GUI}. \\
        \bottomrule
    \end{tabular}
\end{table}
\clearpage



\end{document}